\documentclass[aps,prc,twocolumn,superscriptaddress,floatfix,showpacs,nofootinbib]{revtex4-1}
\usepackage{multirow}
\usepackage{graphicx}
\usepackage{color}
\usepackage{amsmath}
\usepackage{amssymb}
\usepackage[colorlinks=true]{hyperref}
\usepackage{tikz}
\usepackage{etoolbox}
\usepackage{bm}
\usepackage[utf8]{inputenc}
\usepackage{cleveref}
\usepackage{amsmath}
\usepackage{amssymb}
\usepackage{verbatim}
\usepackage{float}
\usepackage[bottom]{footmisc}

\newcommand {\sumc}     {{\tt super-MC}}
\newcommand {\trento}   {{\tt TRENTo}}
\newcommand {\hijing}   {{\tt HIJING}}
\newcommand {\urqmd}    {{\tt UrQMD}}
\newcommand {\vishnu}   {{\tt iEBE-VISHNU}}
\newcommand {\vish}     {{\tt VISHNU}}
\newcommand {\vis}      {{\tt VISH2+1}}
\newcommand {\ipglasma} {IP-Glasma}

\newcommand{\T}{\tilde{T}}

\definecolor{dgreen}{cmyk}{1.,0.,1.,0.4}        
\definecolor{orange}{cmyk}{0.,0.353,1.,0.}    

\newcommand{\x}{\mathbf x}

\begin{document}
\title{
Searching for small droplets of hydrodynamic fluid in proton--proton collisions at the LHC}

\author{Wenbin Zhao}
\affiliation{Department of Physics and State Key Laboratory of Nuclear Physics and Technology, Peking University, Beijing 100871, China}
\affiliation{Collaborative Innovation Center of Quantum Matter, Beijing 100871, China}

\author{You Zhou}
\email{You Zhou: you.zhou@cern.ch}
\affiliation{Niels Bohr Institute, University of Copenhagen, Blegdamsvej 17, 2100 Copenhagen, Denmark}
\author{Koichi Murase}
\affiliation{Center for High Energy Physics, Peking University, Beijing 100871, China}
\author{Huichao Song}
\email{Huichao Song: huichaosong@pku.edu.cn}
\affiliation{Department of Physics and State Key Laboratory of Nuclear Physics and Technology, Peking University, Beijing 100871, China}
\affiliation{Collaborative Innovation Center of Quantum Matter, Beijing 100871, China}
\affiliation{Center for High Energy Physics, Peking University, Beijing 100871, China}
\date{\today}
\begin{abstract}
In this paper, we investigate the hydrodynamic collectivity in high-multiplicity events of proton--proton collisions at $\sqrt{s}=$ 13 TeV, using \vishnu{} hybrid model with three different initial conditions, namely, \hijing, \sumc{} and \trento.  With properly tuned parameters, hydrodynamic simulations with each initial model give reasonable descriptions of the measured two-particle correlations, including the integrated and $p_{\rm T}$-differential flow for all charged and identified hadrons.
However, the hydrodynamic simulations fail to describe the negative value of the four-particle cumulant $c_2^v\{4\}$ as measured in experiments.
We find that the four-particle cumulant $c_2^v\{4\}$ is always positive after hydrodynamic evolutions
even though some of the initial models give a negative cumulant $c_2^\varepsilon\{4\}$ for the initial eccentricity.
Further investigations show that the non-linear response between the elliptic flow $v_2$ and the initial eccentricity $\varepsilon_2$
becomes significant in the small p--p systems.
This leads to a large deviation from linear eccentricity scaling and generates additional flow fluctuations, which results in a positive $c_2^v\{4\}$ even with a negative $c_2^\varepsilon\{4\}$ from the initial state.
We also presented the first hydrodynamic calculations of multi-particle mixed harmonic azimuthal correlations in p--p collisions,
such as normalized asymmetric cumulant $nac_n\{3\}$, normalized Symmetric-Cumulant, $nsc_{2,3}\{4\}$ and $nsc_{2,4}\{4\}$.
Although many qualitative features are reproduced by the hydrodynamic simulations with chosen parameters,
the measured negative $nsc_{2,3}\{4\}$ cannot be reproduced.
The failure of the description of negative $c_2\{4\}$ and $nsc_{2,3}\{4\}$ triggers the question on
whether hydrodynamics with a fundamentally new initial state model could solve this puzzle,
or hydrodynamics itself might not be the appreciated mechanism of the observed collectivity in p--p collisions at the LHC.
\end{abstract}


\pacs{25.75.Ld, 25.75.Gz}

\maketitle

\clearpage

\section{Introduction}
\label{section1}

Ultra-relativistic collisions of heavy ions are intended to create a novel state of matter, the Quark-Gluon Plasma (QGP), and to study its properties.
Extensive measurements of various flow observables performed at the Relativistic Heavy Ion Collider (RHIC) and the Large Hadron Collider (LHC) together with the successful descriptions from hydrodynamic calculations
revealed that the created QGP fireball behaves like a nearly perfect liquid with very small specific shear viscosity~%
\cite{Kolb:2003dz,Adams:2005dq,Adcox:2004mh,Arsene:2004fa,Back:2004je,Gyulassy:2004vg,Gyulassy:2004zy,Muller:2006ee,Jacak:2012dx,Muller:2012zq,Heinz:2013th,Gale:2013da,Shuryak:2014zxa,Song:2017wtw}.
Recently, various striking features of collective expansion have been observed in high-multiplicity events of the small collision systems,
such as p--Au, d--Au, $^3$He--Au at RHIC~\cite{ Aidala:2017ajz,PHENIX:2018lia} and p--p and p--Pb at the LHC~\cite{Li:2012hc,Dusling:2015gta,Nagle:2018nvi}.
These features include the long-range ``double ridge" structures in two-particle azimuthal correlations with a large pseudo-rapidity gap even up to 8 units~\cite{Khachatryan:2010gv, CMS:2012qk,Abelev:2012ola,Khachatryan:2015lva,Aad:2015gqa,Aidala:2017pup,Adare:2018toe,Aad:2013fja,Khachatryan:2015waa,Aaboud:2016yar,Khachatryan:2016txc,ATLAS:2017tqk,Acharya:2019vdf}, the changing signs of the 4-particle cumulants~\cite{Aad:2013fja,Abelev:2014mda,Khachatryan:2015waa,Khachatryan:2016txc,Aaboud:2016yar,Aaboud:2017acw,ATLAS:2017tqk,Aidala:2017ajz,Acharya:2019vdf} and $v_2$ mass ordering of identified hadrons~\cite{ABELEV:2013wsa,Khachatryan:2014jra,Adare:2014keg,Khachatryan:2016txc}, etc.

These observed flow-like signals in the small systems can be quantitatively or semi-quantitatively described
by hydrodynamic calculations~\cite{ Bozek:2011if,Bzdak:2013zma,Qin:2013bha,
Nagle:2013lja,Werner:2013tya,Werner:2013ipa,Bozek:2013ska,Schenke:2014zha,Bozek:2014cya,Bozek:2015swa,
Shen:2016zpp,Weller:2017tsr,Mantysaari:2017cni,Zhao:2017rgg}, which translate initial
spatial anisotropies into final momentum anisotropies of produced hadrons with the collective
expansion of the bulk matter. Besides, other model calculations based on final state interactions, such as transport models~\cite{Bzdak:2014dia,Kurkela:2018ygx,Nie:2018xog,Sun:2019gxg,Wei:2019wdt}, hadronic rescatterings~\cite{Zhou:2015iba,Romatschke:2015dha}, a string rope and shoving mechanism~\cite{Bierlich:2017vhg} have also been performed to study the collective behavior of small systems. Alternatively, the color glass condensate (CGC) or IP-Plasma focused on initial state effects~\cite{Dusling:2012cg,Dusling:2012iga,Kovner:2012jm,Kovchegov:2012nd,
Lappi:2015vta,Schenke:2015aqa,Schenke:2016lrs,Dusling:2017dqg,Dusling:2017aot,Mace:2018yvl} can also qualitatively reproduce many features of collectivity. The origin of the observed collective behavior in the small systems is still under intense debate.
Recently, the model calculations in~\cite{Zhao:2019ehg} showed that the quark coalescence procedure is necessary to reproduce the number of constituent quark scaling of $v_2$ at intermediate $p_T$ in high-multiplicity  p--Pb collisions at $\sqrt{s}=$13 TeV, which demonstrate the importance of the partonic degrees of freedom and possible formation of QGP in the small p--Pb systems.

Recently, the collectivity and possible formation of QGP in high-multiplicity p--p collisions at the LHC energies has also attracted lots of attention. Compared to larger collision systems, the corresponding non-flow contributions, such as mini-jets or resonance decays, become more significant. In the measurements of two-particle correlations, two different non-flow subtraction methods, template fit \cite{Aad:2015gqa,Aaboud:2016yar,ATLAS:2017tqk} and peripheral subtraction \cite{Khachatryan:2016txc}, have been applied to remove the non-flow contaminations for the extracted  flow harmonics.
Meanwhile, multi-particle cumulants have been systematically measured, which provide more insights for the collective phenomenon in high-multiplicity p--p collisions. Compared to the two-particle correlations, multi-particle cumulants, by construction, have the advantage of suppressing short-range two-particle correlations~\cite{Voloshin:2008dg,Jia:2017hbm}. Besides, two- and three-subevent methods have been implemented to further suppress the remaining non-flow contaminations, which are also much less sensitive to the multiplicity fluctuations compared to the standard method~\cite{Jia:2017hbm,Aaboud:2017blb,Aaboud:2018syf}. It was found that $c_2\{4\}$ turns to negative value in high-multiplicity events of p--p collisions, which gives the real value of the flow coefficients $v_{2}\{4\}$ through the relation $c_2\{4\} = - v_{2}\{4\}^4$ and strongly indicates the existence of anisotropic flow in the small p--p systems~\cite{Aaboud:2017blb,Aaboud:2018syf}. Furthermore, ALICE~\cite{Acharya:2019vdf}, ATLAS~\cite{Aaboud:2018syf} and CMS~\cite{Sirunyan:2017uyl} have measured the correlations between different flow harmonics $v_n$ and $v_m$ via three- or four-particle cumulants in p--p collisions, which shows  negative correlations between $v_2$ and $v_3$ and  positive correlations between $v_2$ and $v_4$ with similar relative correlation strengths as measured in p--Pb and Pb--Pb systems. It is thus on-time and important to investigate these collective flow signatures by hydrodynamic models and to discuss whether hydrodynamic calculations could describe two- and multi-particle cumulants simultaneously in the small p--p systems created at the LHC.


In our previous work~\cite{Zhao:2017yhj}, we found that, with properly tuned parameters, hydrodynamic simulations with {\tt HIJING} initial conditions can nicely describe the two-particle correlations in p--p collisions at $\sqrt{s}=$13 TeV, including the integrated $v_2\{2\}$, differential elliptic flow $v_2({p_T})$ for all charged hadrons and for identified particles ($K^0_S$ and $\Lambda$). However, the measured negative $c_2\{4\}$, which has been usually interpreted as evidence of hydrodynamic flow, could not be reproduced by our hydrodynamic calculations which showed a positive value of $c_2\{4\}$. It is still unknown if the wrong sign of $c_2\{4\}$ is due to the incorrect initial conditions from {\tt HIJING} or due to the application of the hydrodynamic model to p--p collisions.

To address these questions, in this paper, we implement three different initial conditions, called \hijing~\cite{Zhao:2017rgg}, \sumc~\cite{Welsh:2016siu} and \trento~\cite{Moreland:2018gsh}, to the \vishnu{} hybrid model simulations to study various flow observables in p--p collisions, especially on the four-particle cumulant $c_2\{4\}$ and mixed harmonic three- and four-particle azimuthal correlations. To better understand the non-linear hydrodynamic evolution in the small systems, we also investigate the response between the initial $\varepsilon_2$ and final $v_2$.  In addition, we study the effects of pre-equilibrium dynamics in the p--p collision by including the free-streaming evolution before the hydrodynamic simulations.


This paper is organized as follows: in the next section, we will give an introduction of the \vishnu{} hydrodynamic model and the initial conditions of \hijing{}, \sumc{} and \trento{} and explain the setups. Section~\ref{sec:result} presents the model calculations, the comparison to the experimental data, and the related discussion. Section~\ref{sec:summary} gives a brief summary of this paper.


\section{The model and set-ups}
\label{sec:model}
\subsection{\vishnu \ hybrid model}

\vishnu~\cite{Shen:2014vra} is an event-by-event version of hybrid model \vish~\cite{Song:2010aq} that combines 2+1D viscous hydrodynamics \vis~\cite{Song:2007fn,Song:2007ux} to describe the QGP expansion with a hadron cascades model \urqmd~\cite{Bass:1998ca,Bleicher:1999xi} to simulate the evolution of hadronic matter.  Based on the Israel--Stewart formalism, \vis\ solves the transport equations for the energy-momentum tensor $T^{\mu \nu}$ and shear stress tensor $\pi^{\mu \nu}$ with a state-of-the-art equation of state (EoS) s95-PCE~\cite{Huovinen:2009yb,Shen:2010uy} as an input to simulate viscous fluid expansion of the hot QCD matter with longitudinal boost-invariance. For simplicity, we neglect the bulk viscosity, net baryon density, and heat conductivity and assume a constant specific shear viscosity $\eta/s$. The hydrodynamic evolution matches the hadron cascade simulations at a switching temperature $T_{\rm sw}$, where various hadrons are emitted from the switching hyper-surface for the succeeding \urqmd\ evolution.

To systematically investigate the hydrodynamic collectivity of p--p system and its dependence on the initial condition models, we implement three different initial condition model, namely, modified \hijing~\cite{Zhao:2017rgg}, \sumc~\cite{Welsh:2016siu} and \trento~\cite{Moreland:2018gsh}. In general, these three initial conditions neglect the pre-equilibrium dynamics and set the initial flow velocity and the shear-stress tensor to be zero for the succeeding hydrodynamic simulations, which are also the default settings of our calculations. In this paper, for one parameter set of \trento\ initial condition, we prepared a version including the free-streaming evolution before hydrodynamics to study the pre-equilibrium effects. Below is a brief description of these three initial condition models.

\subsection{\hijing\ initial condition}
In \hijing\ ~\cite{Wang:1991hta, Deng:2010mv, Deng:2010xg},  the radial density  of the colliding protons is the Woods--Saxon shapes, and the produced jet pairs and excited nucleus are treated as independent strings, where the hard jet productions are calculated by  pQCD, and the soft interactions are treated as gluon exchange within Lund string model. For the \hijing\ initial condition developed in Ref.~\cite{Zhao:2017rgg}, it assumes that the mother strings that break into independent partons quickly form several hot spots for the succeeding hydrodynamic evolution. The center positions of these mother strings $(x_c, y_c)$ are sampled by the Woods--Saxon distribution, and the positions of the produced partons $(x_i, y_i)$ within the strings are sampled with a Gaussian distribution with a width $\sigma_R$: $\exp[-\frac{(x_i-x_{c})^{2}+(y_i-y_{c})^{2}}{2\sigma_{R}^2}]$.

The initial energy density profiles in the transverse plane for the 2+1D hydrodynamic evolution are constructed from the energy depositions of emitted partons, together  with an additional Gaussian smearing~\cite{Zhao:2017rgg}
\begin{equation}
	e(x,y) = K\sum_{i}\frac{p_{i}U_{0}}{2\pi\sigma_0^{2}\tau_{0}\Delta\eta_{s}}\exp\left[-\frac{(x-x_{i})^{2}+(y-y_{i})^{2}}{2\sigma_0^2}\right], \label{eq:epsilon}
\end{equation}
where $\sigma_0$ is the Gaussian smearing factor, $p_{i}$ is the momentum of the produced parton $i$,
and $K$ is an additional normalization factor. Here, we neglect the initial flow $U_{0}$ and only consider the  partons within the mid-rapidity $|\eta|<1$ (for related details, please also refer to Ref.~\cite{Zhao:2017rgg}).

\subsection{\sumc\ initial condition}
For p--p collisions,  \sumc\ model with sub-nucleonic  fluctuations~\cite{Welsh:2016siu} assumes the colliding protons consist of three valence quarks, and  the collisions between valence quarks
depose a fraction of the kinetic energy of the colliding systems into the initial energy of the newly formed matter, which fluctuates from event to event.  Following Ref.~\cite{Welsh:2016siu}, the initial entropy density of the produced matter is modeled as
\begin{equation}
\label{eq26}
   s(\mathbf{r}) = \frac{\kappa_s}{\tau_0} \sum_{k=1}^3 \frac{\gamma_k^{(i)}}{2\pi\sigma_g^2}
   \exp\biggl[-\frac{(\mathbf{r}-\mathbf{r}_k^{(i)})^2}{2\sigma_g^2}\biggr],
\end{equation}
where $\gamma_k^{(i)}$  ($i=1,2,3$) are the random weighting factors that used to fit the multiplicity distributions in p--p collisions, ${\bf{r}}_i$ ($i=1,2,3$)  is the position of three valence quarks which is  distributed  according to a Gaussian probability distribution. $\sigma_g$ is a factor to describe the shape of quark density distribution together with a consideration of low-$x$ gluon contributions~\cite{Welsh:2016siu}.

\begin{table}[t]
\centering  %
\caption{Four parameter sets of  \vishnu{} simulations  with \hijing{} initial condition for p--p collisions at $\sqrt{s}=$ 13 TeV. }
         \label{tb:hijing}
\begin{tabular}{|l|c|c|c|c|c|c|}
\hline
   &$\sigma_R$&$\sigma_{0}$&$\tau_0$ & $\eta/s$  &$T_{\rm sw}$(MeV) \\ \hline %
Para-I  &1.0&0.4&0.1&0.07 &147\\        \hline  %
Para-II &0.8&0.4&0.2&0.08&148\\        \hline  %
Para-III &0.4&0.2&0.6&0.20   &148\\        \hline  %
Para-IV &0.6&0.4&0.4&0.05   &147\\        \hline  %
\end{tabular}]
\centering  %
\caption{Three parameter sets of \vishnu{} simulations with \sumc{} initial condition for p--p collisions at $\sqrt{s}=$ 13 TeV. }
         \label{tb:supermc}
\begin{tabular}{|l|c|c|c|c|c|c|}
\hline
   &$\sigma_g$&$\tau_0$ & $\eta/s$  &$T_{\rm sw}$(MeV) \\ \hline %
Para-I  &0.4&0.8&0.22 &148\\        \hline  %
Para-II &0.5&0.6&0.12  &149\\        \hline  %
Para-III &0.5&0.8&0.16   &148\\        \hline  %
\end{tabular}
\centering  %
\caption{Three parameter sets of \vishnu{} simulations  with \trento{} initial condition for p--p collisions at $\sqrt{s}=$ 13 TeV. }
         \label{tb:trento}
\begin{tabular}{|l|c|c|c|c|c|c|c|c|}
\hline
   &$p$&$v$&$k$&$n_c$&$\tau_0$ & $\eta/s$ &$T_{\rm sw}$(MeV) \\ \hline   %
Para-I  &0.5&0.3&1.5&4&0.2&0.08&149\\        \hline  %
Para-II  &0.0&0.2&0.81&6&0.6&0.28&149\\        \hline  %
Para-III  &0.5&0.2&1.0&4&0.8&0.28&149\\        \hline  %
\end{tabular}
\end{table}

  \begin{figure*}[thb]
\begin{center}
\includegraphics[width=0.95\textwidth]{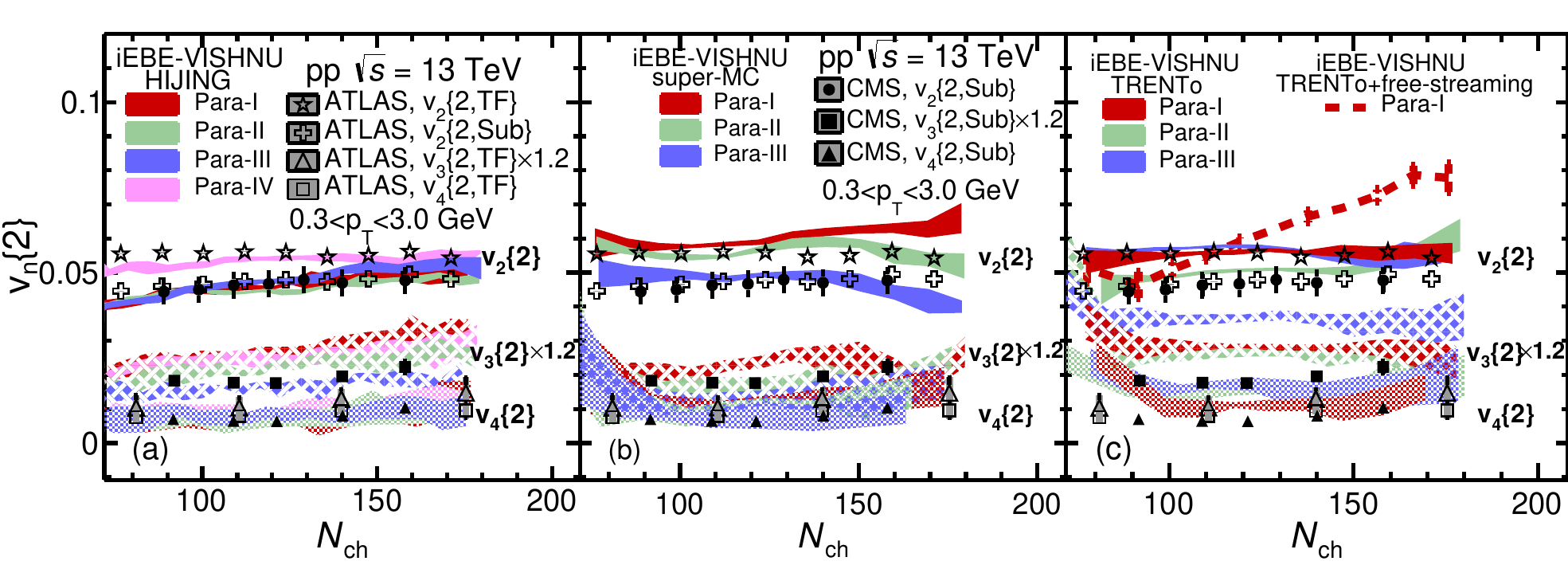}
\caption{(Color online)  $v_2\{2\}$,  $v_3\{2\}$ and  $v_4\{2\}$  as a function of $N_{\rm ch}$  in  p--p collisions at $\sqrt{s}=$ 13 TeV, calculated by \vishnu{} with \hijing{} (a), \sumc{} (b) and \trento{} (c) initial conditions. The CMS and ATLAS  data are taken from Refs.~\cite{Khachatryan:2016txc,Sirunyan:2017uyl} and Refs.~\cite{ATLAS:2017tqk,Aaboud:2018syf}, respectively.}
\label{fig:v22}
\end{center}
\end{figure*}

 \subsection{\trento\ initial condition}
\trento\ is a parameterized initial condition model, which generates the initial entropy density via the reduced thickness function~\cite{Moreland:2014oya,Bernhard:2016tnd}:
\begin{equation}
  s = s_0 \left( \frac{\T_A^p + \T_B^p}{2} \right)^{1/p},
  \label{eq:entropy}
\end{equation}
where $\T(x, y)$ is the modified participant thickness function, $s_0$ is a normalization factor, and $p$ is a tunable parameter which makes \trento\ model effectively interpolates among different entropy deposition schemes, such as {\tt{KLN}}, {\tt{EKRT}}, {\tt WN}, etc.~\cite{Moreland:2014oya,Bernhard:2016tnd,Moreland:2018gsh}.

For proton--proton collisions, \trento\ is modified with the sub-nucleonic structure~\cite{Moreland:2018gsh} so that $\T(x, y)$ is written as $\T(x, y)\equiv \int dz\, \frac{1}{n_c} \sum_{i=1}^{n_c} \gamma_i\, \rho_c \,(\mathbf{x} - \mathbf{x}_i \pm \mathbf{b}/2)$, where $n_c$ is the number of independent constituents in a proton, $ \gamma_i$ ($i=1,2, ..., n_c$) is a random weighting factor with the unit mean and variance $1/k$,  $\mathbf{x}_i$ ($i=1,2, ..., n_c$) are the positions of constituents, $\bf{b}$ is the impact parameter,  and $\rho_c$ is the density of constituents written in a Gaussian form: $ \rho_c(\mathbf{x}) = \frac{1}{(2\pi v^2)^{3/2}} \exp(-\frac{\x^2}{2 v^2})$, and $v$ is a tunable effective width of nucleons. \\[-0.10in]

\noindent\underline{\emph{\trento{} initial condition with free streaming:}} \\[-0.10in]

For \trento{} initial condition, we also construct another type of the initial condition with free-streaming to include the effects of pre-equilibrium dynamics before hydrodynamic evolution.  Following~Refs.~\cite{Broniowski:2008qk,Liu:2015nwa,Moreland:2018gsh},  we assume the  particle density of non-interacting massless particles at the very beginning is proportional to entropy density described by Eq.~(\ref{eq:entropy}), and
then free streaming these massless partons till the proper time $\tau_0$ to obtain the boost-invariant  energy-momentum tensor $T^{\mu \nu}(x,y, \tau_0)$. After that, we implement the following Landau matching condition to obtain the initial energy density $e(x,y, \tau_0)$ and fluid velocity $u^\mu(x,y, \tau_0)$:
\begin{equation}
  T^{\mu\nu} u_\nu = e u^\mu ,
\end{equation}
and the initial shear stress tensor and bulk pressure can be calculated with:
\begin{align}
  \Pi &= -\frac{1}{3} \Delta_{\mu\nu} T^{\mu\nu} - P,\\
  \pi^{\mu\nu} &= T^{\mu\nu} - e u^\mu u^\nu + (P + \Pi) \Delta^{\mu\nu},
\end{align}
with the spatial projector being $\Delta^{\mu\nu} = g^{\mu\nu} - u^\mu u^\nu$,
together with an equation of state of  $P=\frac{1}{3} e$  for the massless ideal gas at the initial state.\\ [0.10in]



For p--p collisions at $\sqrt{s}=$13 TeV, we implemented several sets of parameters for each of these three or four different initial conditions. These parameters are roughly tuned to approximately fit the $p_T$-spectra~\cite{ALICE13teV} and $v_2\{2\}$~\cite{Khachatryan:2016txc,Sirunyan:2017uyl,ATLAS:2017tqk,Aaboud:2018syf} measured in experiments. Note that these data are not enough to fully constrain the free parameters in hydrodynamic simulations. Since this paper is aimed to investigate the sign of $c_2\{4\}$,   mixed harmonic three- and four-particle
azimuthal correlations, and the effects of non-linear evolution rather than make quantitative descriptions and prediction for p--p collisions, we chose three or four sets of parameters for each initial condition, as listed in Tables~\ref{tb:hijing},~\ref{tb:supermc} and~\ref{tb:trento}. For \trento{} initial condition with the parameter set Para-I, we consider two cases with and without free-streaming as described above.

\section{Results and Discussions}
\label{sec:result}

\subsection{2-particle cumulant}

  \begin{figure*}[thb]
\begin{center}
\includegraphics[width=0.95\textwidth]{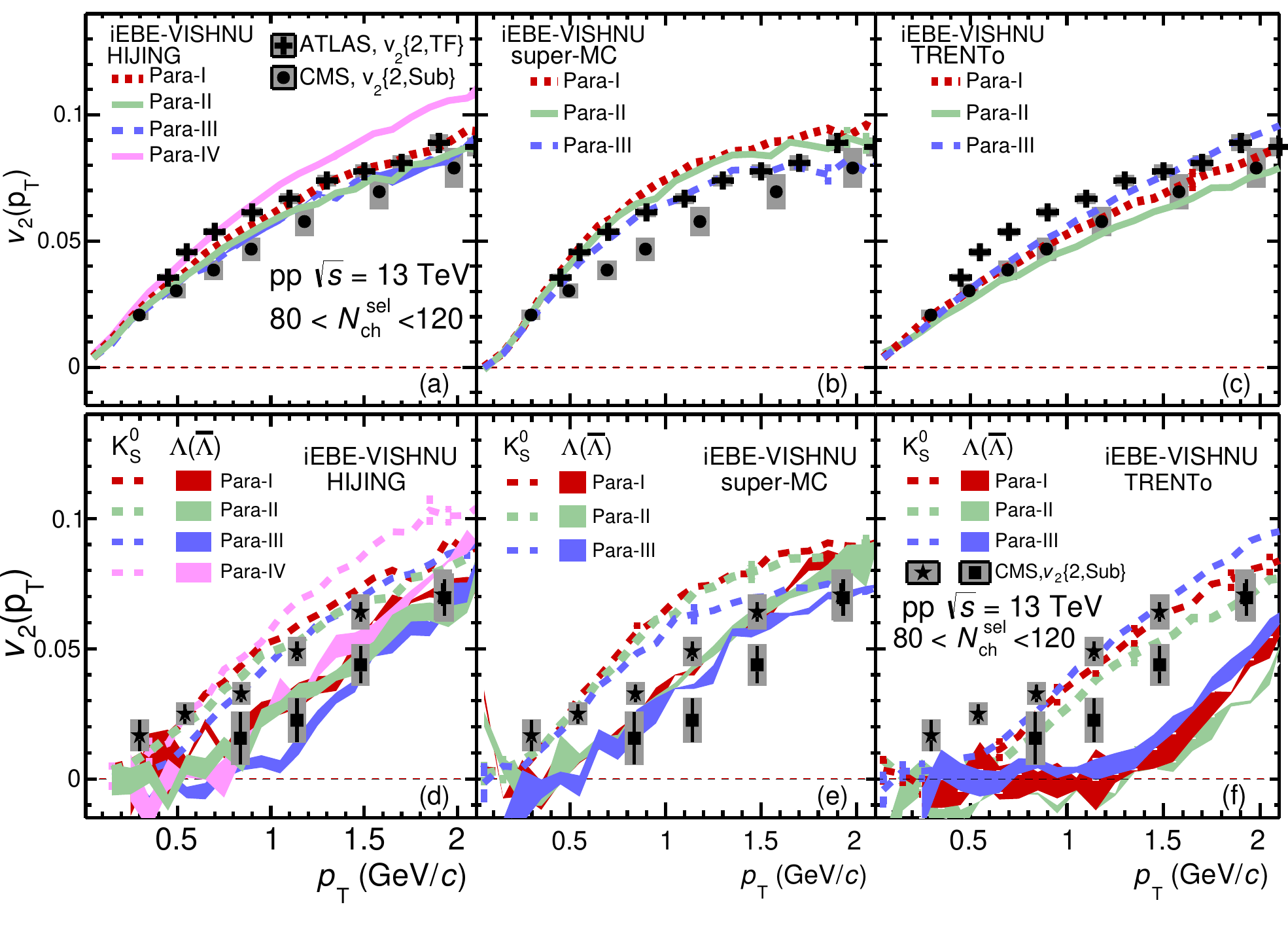}
\caption{(Color online)  $v_2(p_T)$ for all charged hadrons  (a)--(c),  for   $K_S^0$ and $\Lambda$ (d)--(f) in  p--p collisions at $\sqrt{s}=$ 13 TeV, calculated by \vishnu{} with \hijing, \sumc{} and \trento{} initial conditions. The CMS and ATLAS  data are taken from Refs.~\cite{Khachatryan:2016txc} and \cite{Aaboud:2016yar}, respectively.}
\label{fig:v2ptall}
\end{center}
\end{figure*}

  \begin{figure*}[thb]
\begin{center}
\includegraphics[width=0.95\textwidth]{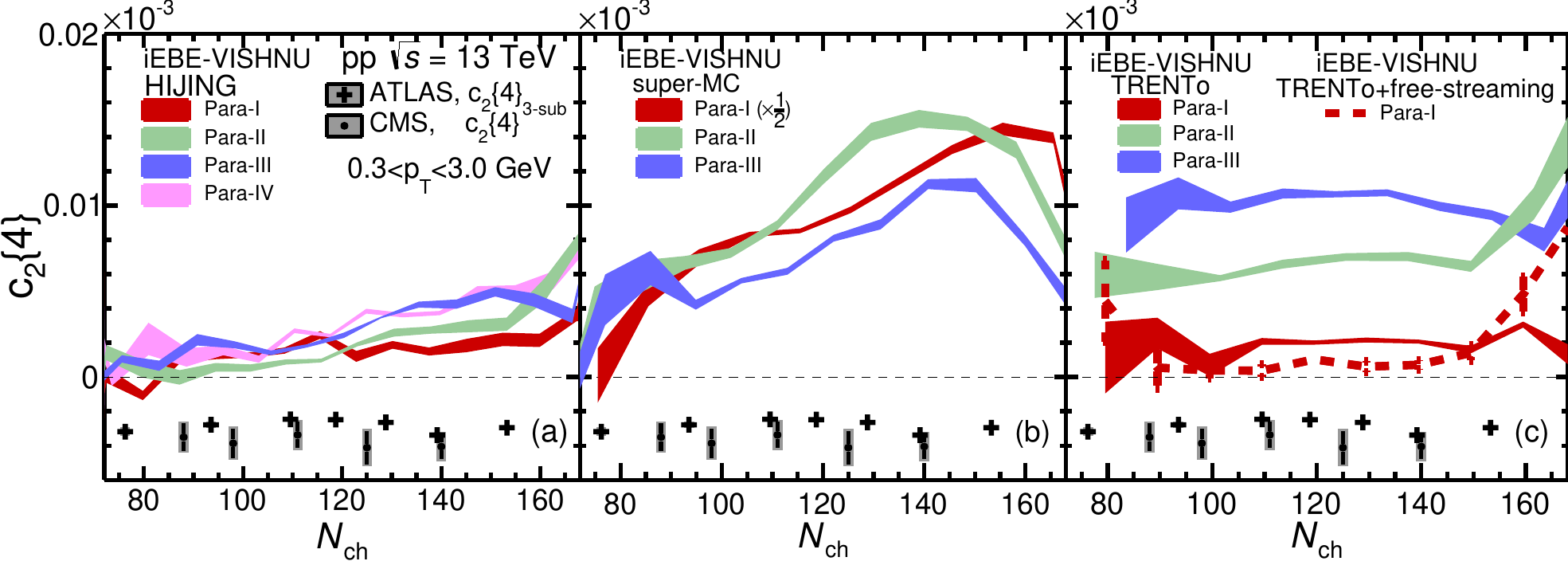}
\caption{(Color online)  $c_2\{4\}$  as a function of $N_{\rm ch}$   in p--p collisions at $\sqrt{s}=$ 13 TeV, calculated by  \vishnu{} with \hijing{} (a), \sumc{} (b) and \trento{} (c) initial conditions using standard cumulant method. The CMS data with standard cumulant method and the ATLAS data with three-subevent method are taken from Refs.~\cite{Khachatryan:2016txc} and~\cite{ATLAS:2017tqk}, respectively.}
\label{fig:c24}
\end{center}
\end{figure*}

\begin{figure*}[t]
\begin{center}
\includegraphics[width=0.95\textwidth]{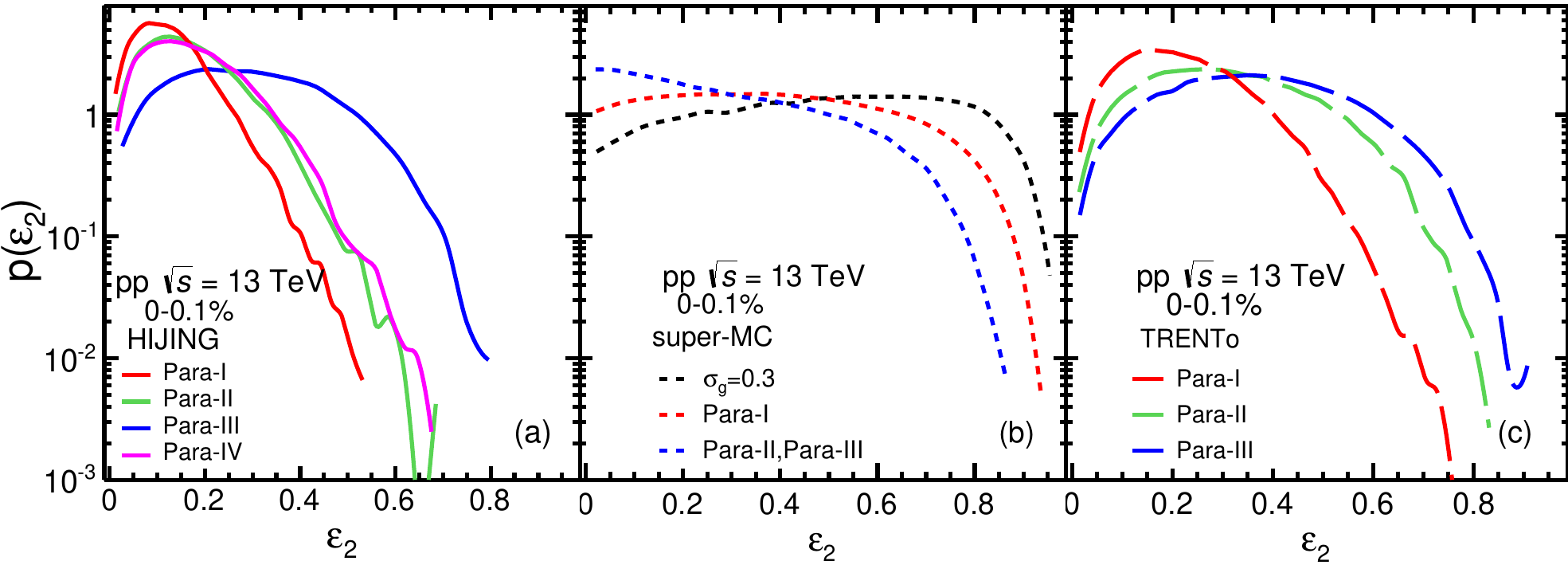}
\caption{(Color online)  Event-by-event  $\varepsilon_2$ distributions $P(\varepsilon_2)$ of \hijing{} (a), \sumc{} (b) and \trento{} (c) initial conditions at 0--0.1\% centrality bin in p--p collisions at $\sqrt{s}=13\ \text{TeV}$. }
\label{fig:pecc}
\end{center}
\end{figure*}

\begin{figure*}[t]
\begin{center}
\includegraphics[width=0.35\textwidth]{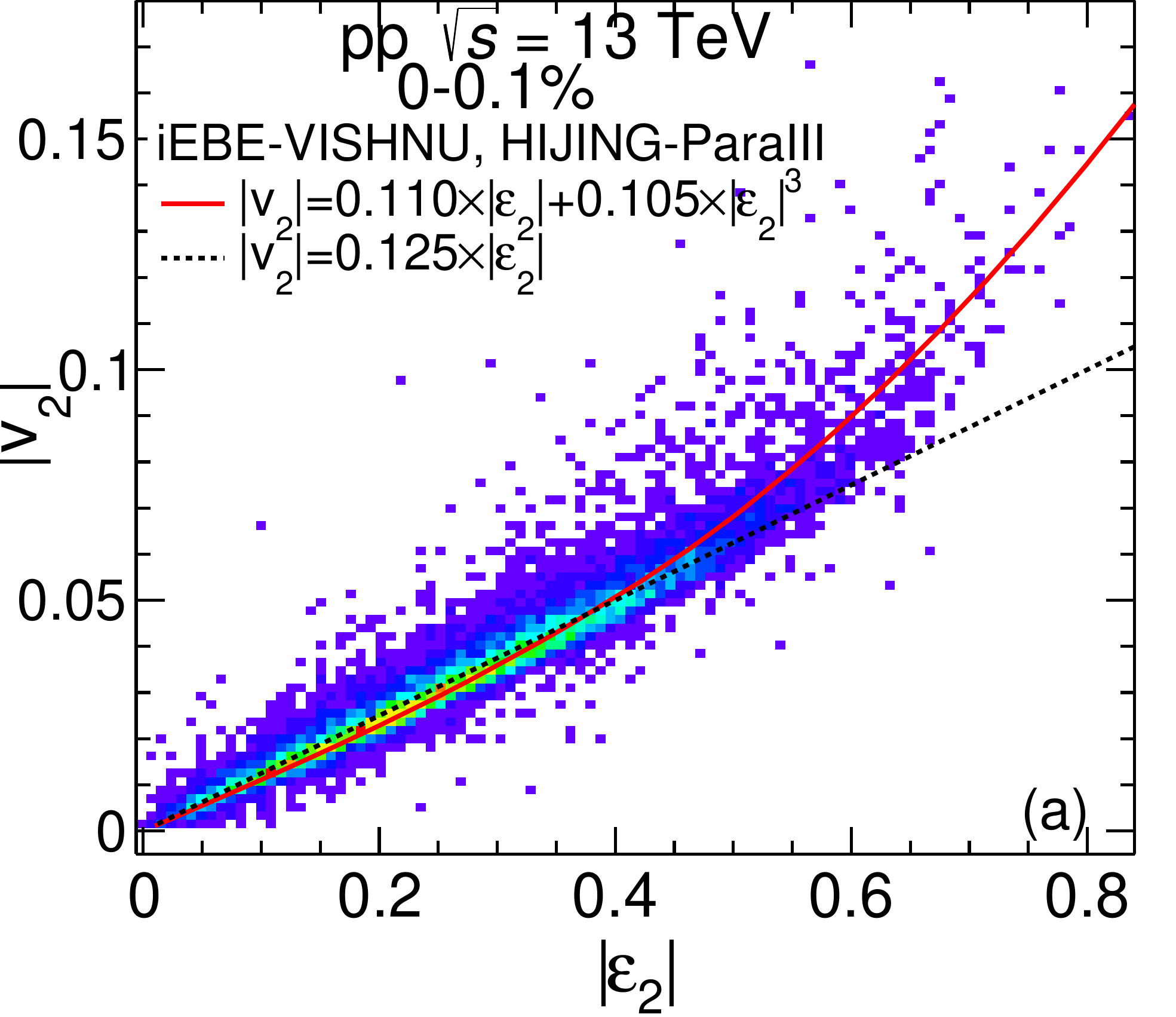}\includegraphics[width=0.35\textwidth]{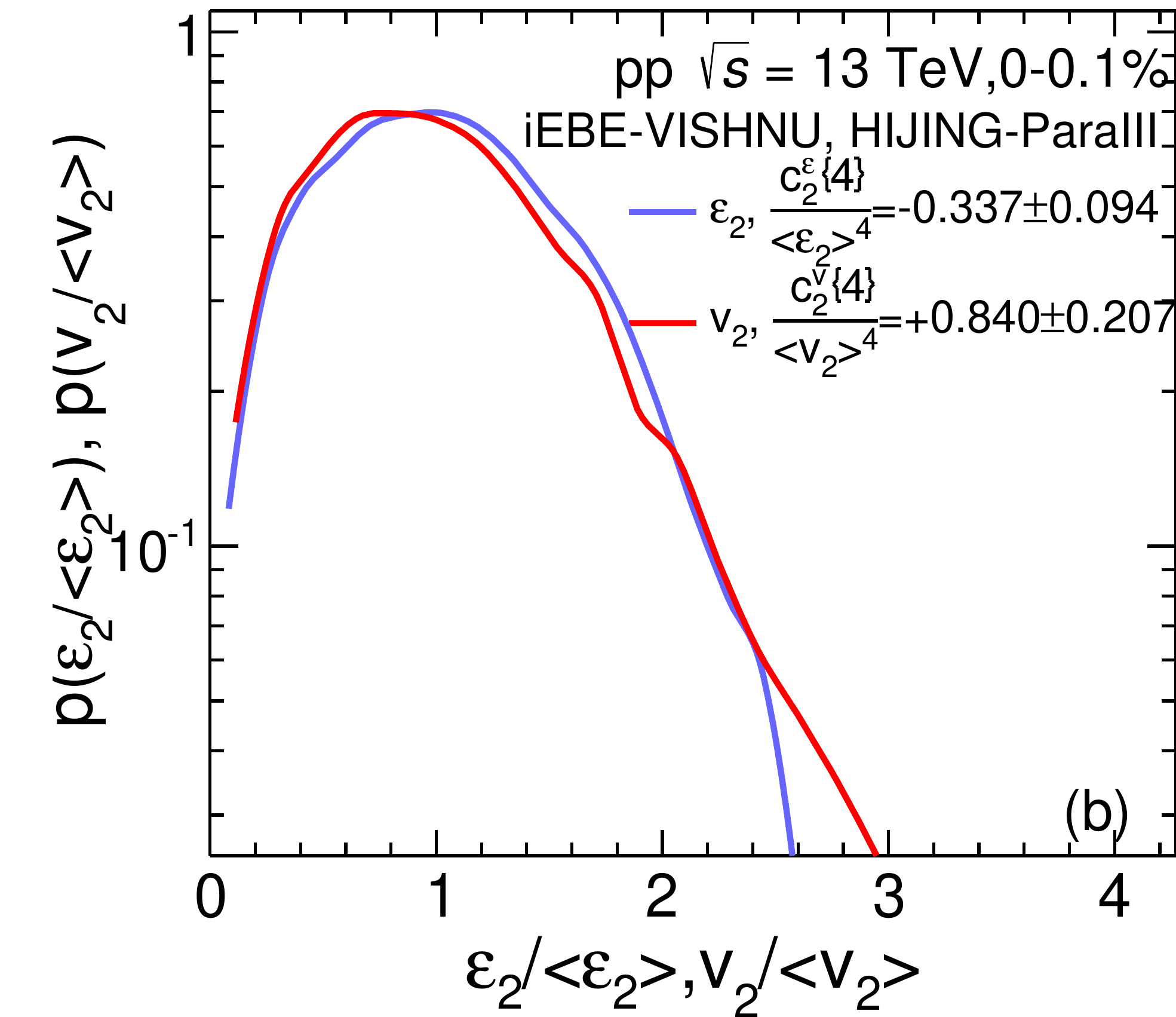}
\caption{(Color online) Left panel: the scatter points between the $v_2$ and $\varepsilon_2$, together with a linear fitting and a non-linear fitting with both linear and cubic terms. Right panel: the comparison between the scaled event-by-event $\varepsilon_2$ distribution and scaled $v_2$ distributions for \vishnu{} simulations with \hijing{} initial condition (Para-III) at 0--0.1\% p--p at $\sqrt{s}=$ 13 TeV.  }
\label{fig:pv2e2}
\end{center}
\end{figure*}

With various sets of parameters for these three initial conditions, \hijing{}, \sumc\ and \trento{}, as listed in Tables~\ref{tb:hijing}, \ref{tb:supermc} and \ref{tb:trento}, we calculate the integrated $v_n\{2\}$ ($n=2$, $3$ and $4$) as a function of multiplicity for p--p collisions at $\sqrt{s}=$ 13 TeV, using \vishnu{}  together with an application of the two-subevent method with the pseudorapidity gap $|\Delta\eta| > 0$, kinematic cuts $0.3 < p_{\rm T} < 3.0 $\ GeV/$c$ and $|\eta|<2.4$.  To eliminate the effects of multiplicity fluctuations, we implement the same method as used in experimental analysis and in our early paper~\cite{ATLAS:2017tqk,Zhao:2017rgg}, which first obtain the 2- and 4-particle cumulants  within the multiplicity class with the number of charged hadrons  $N_{\rm ch}^{\rm sel}$ with $0.3 < p_{\rm T} < 3.0 $\ GeV/$c$ and $|\eta|<2.4$, and then map it to the  number of charged hadrons $N_{\rm ch}$ with  $0.4 < p_{\rm T} $\ GeV/$c$ and $|\eta|<2.4$ to compare with the experimental data. Fig.~\ref{fig:v22} presents the comparison between our hydrodynamic calculations and the experimental measurements from ATLAS~\cite{Khachatryan:2016txc,Sirunyan:2017uyl} and CMS~\cite{ATLAS:2017tqk,Aaboud:2018syf}. It shows that hydrodynamic simulations with these three different initial conditions can generally reproduce the multiplicity dependence of the integrated $v_2\{2\}$  as we could expect from tuning the related parameters.
Note that these four sets of parameters, Para-I--IV, in \hijing\ initial condition, are the same as  we used in Ref.~\cite{Zhao:2017rgg}, which are tuned to fit $v_2\{2\}$ data obtained from the ``peripheral subtraction'' method (Para-I--III) and  from the ``template fit'' method (Para-IV), respectively. For \sumc\ and \trento\ initial conditions, we choose one set of parameters (Para-III for \sumc{} and Para-II for \trento{}) to describe the $v_2 \{ 2 \}$ data with ``peripheral subtraction'' method, and the other parameter sets  to approximately describe the data with ``template fit'' method. In general, hydrodynamic calculations approximately describe  $v_{4}\{2\}$ from CMS and ATLAS, but tend to overestimate the measured $v_{3}\{2\}$ with both ``peripheral subtraction'' and ``template fit'' methods, especially for the ones obtained with \trento\ initial conditions. On the other hand, $v_{3}\{\rm 2\}$ data  from ``peripheral subtraction'' and ``template fit'' methods also largely deviate from each other, and it is still under debate on which method gives a better non-flow subtraction for the odd flow harmonics~\cite{Aaboud:2018syf}.

For the parameter set of Para-I of \trento\ initial condition, we also include the pre-equilibrium evolution with an infinitely weak coupling limit (dashed red line), which free-streams the initial state to proper time $\tau_0$ before instantaneously switching to hydrodynamic simulations~\cite{Liu:2015nwa,Moreland:2018gsh}. It shows that such pre-equilibrium dynamics not only affects the magnitude of $v_2\{2\}$ but also affects its dependence on the multiplicity, which seems excluded by the experimental data.

From the hydrodynamic calculations shown in Fig.~\ref{fig:v22}, it is clear that the flow coefficients of $v_2$, $v_3$ and $v_4$ in p--p collisions could provide certain constraints on the parameter settings for model calculations with various initial conditions. A simultaneous description of $v_2$, $v_3$ and $v_4$ is one of essential steps to validate the applicability of hydrodynamic simulations in small systems.

In Fig.~\ref{fig:v2ptall}, we calculate differential elliptic flow $v_2(p_T)$ for all charged hadrons (a)--(c) and  for $K_S^0$ and $\Lambda$ (d)--(f) for the multiplicity range $80< N_{\rm ch}^{\rm sel} <120$ with the two-particle cumulant method with a pseudorapidity gap $|\Delta\eta|>0$. \vishnu\ with \hijing, \sumc\ or \trento\ initial conditions can roughly describe $v_2(p_T)$ for all charged hadrons measured from CMS and ATLAS with the ``peripheral subtraction'' or ``template fit'' method. More specifically, as mentioned previously in Ref.~\cite{Zhao:2017rgg}, the calculations of Para-I, II and III of \hijing\ initial condition, which are tuned for ``peripheral subtraction'', give a satisfactory description of the data. In contrast, Para-IV of \hijing\ initial condition tuned for ``template fit'' slightly overpredicts the data above 1.0\ GeV/$c$. For \sumc\ and \trento\ initial conditions, hydrodynamic calculations can roughly describe $v_2(p_T)$ data  for all charged hadrons.

The panels (d)--(f) of Fig.~\ref{fig:v2ptall} present $v_2(p_T)$ for identified hadrons, which show clear $v_2$ mass ordering between $K_S^0$ and $\Lambda$ for both CMS measurements and our \vishnu\ calculations. The hydrodynamic predictions with \hijing\ initial condition (Para-I and II) can nicely describe the data. However, the calculations with \sumc\ initial condition tend to overestimate  $v_2(p_T)$ of $K_s^0$ and $\Lambda$, and  the calculations  with \trento\ initial condition tend to  underestimates the data of $\Lambda$. The mass splitting between $K_S^0$ and $\Lambda$ is more significant for calculations with \trento\ initial condition. Such larger mass splitting of $v_2$  indicates a stronger radial flow development during the hydrodynamic evolution. This is consistent with what we have seen (but not shown here) in the $p_T$-spectra (the spectra obtained with \trento\ initial condition is harder than the others~\cite{Provitenote}), which can also provide certain constrains on the initial conditions.

\subsection{4-particle cumulant}
In Fig.~\ref{fig:c24}, we study the four-particle cumulants of the second harmonics, $c_2\{4\}$, in high-multiplicity proton--proton collisions at $\sqrt{s} = 13\ \text{TeV}$. Although \vishnu\ can roughly describe the measured $v_n\{2\}$ using these three initial conditions with the properly tuned parameters, the predicted  $c_2\{4\}$ are always positive in the high-multiplicity region and fail to reproduce the negative $c_2\{4\}$ as measured in experiments. In Ref.~\cite{Zhao:2017rgg} we have found that the positive $c_2\{4\}$ from hydrodynamic simulations with {\tt HIJING} initial condition is not due to the effects of non-flow contributions or multiplicity fluctuations. We also demonstrated that    the standard method, two-subevent method and three-subevent method almost give the same value of $c_2\{4\}$ for such flow-dominated systems.  The panels (b) and (c) in Fig.~\ref{fig:c24} also show, for the two newly implemented \sumc{} and \trento{} initial conditions,  \vishnu{} still generates a positive $c^v_2\{4\}$ even for these parameter sets associated with a negative $c^\varepsilon_2\{4\}$ for 0--0.1\% events in the initial states, as listed in Table~\ref{tb:c24initial}. Note that recent {\tt MUSIC} hydrodynamic simulations with \ipglasma{} initial conditions also give positive $c^v_2\{4\}$ for the entire multiplicity range in p--p collisions at $\sqrt{s}=13\ \text{TeV}$~\cite{BjoernQM}. We thus emphasize that hydrodynamic simulations do not necessarily produce negative $c^v_2\{4\}$, and the observed negative $c^v_2\{4\}$ in experiments does not necessarily suggest hydrodynamic flow in small systems.

\begin{table}[t]
\centering  %
\caption{$c^{\varepsilon}_2\{4\}$ for 0--0.1\% centrality calculated by \hijing, \sumc, and \trento\ initial conditions with three or four sets of parameters. }
         \label{tb:c24initial}
\begin{tabular}{|l|c|c|c|c||}
\hline
   &\hijing{} ($\times 10^{-4}$)&\sumc{} ($\times 10^{-4}$)&\trento{} ($\times 10^{-4}$)  \\ \hline   %
Para-I  &$2.5\pm0.5$& $- 32.0\pm3.3$&$- 0.64\pm0.03$\\        \hline  %
Para-II  &$3.2\pm1.3$& $50.0\pm1.0$&$- 30.7\pm0.54$\\        \hline  %
Para-III  &$- 22.0\pm6.0$&$50.0\pm1.0$&$- 92.4\pm1.5$\\        \hline  %
Para-IV  &$3.4\pm1.2$& & \\        \hline  %
\end{tabular}
\end{table}

With such findings, we then focus on the effects of non-linear hydrodynamic evolution on the four-particle cumulant $c_2\{4\}$. Specifically, if the final $v_2$ has a linear response to the initial $\varepsilon_2$, the scaled $v_2$ distributions $P(v_2/\langle v_2\rangle)$ should overlap with the scaled $\varepsilon_2$ distributions $P(\varepsilon_2/\langle\varepsilon_2\rangle)$, which is the case for central and mid-central Pb--Pb collisions~\cite{Gale:2012rq}. If such a linear response holds in p--p collisions, the final state $c^v_2\{4\}$ is expected to have the same sign as the initial state $c_2^{\varepsilon}\{4\}$. However, hydrodynamic simulations did not confirm such such expectation. As shown in Figs.~\ref{fig:c24} and \ref{fig:pecc}, and Table~\ref{tb:c24initial}, the negative initial $c^{\varepsilon}_2\{4\}$ (e.g., Para-III of \hijing{}, Para-I of \sumc{}, and Para-I--III of \trento{}) still lead to a positive $c^v_2\{4\}$ at final state after the hydrodynamic evolution. We also find that even though the $c^{\varepsilon}_2\{4\}$ of Para-III of \trento{} initial conditions is more negative than that of Para-I in Table~\ref{tb:c24initial},  the initial $\varepsilon_2$ distribution of Para-III of \trento{} initial condition is ``wider" with larger mean value of $\varepsilon_2$.  The corresponding larger non-linear effects during the evolution lead to a larger positive value of $c_2^v\{4\}$ than the one associated with Para-I\@.  As demonstrated by Figs.~\ref{fig:c24} and \ref{fig:pecc} and also confirmed by additional calculations which are not shown in this paper, similar situations also happen for \vishnu{} simulations with \hijing\ or  \sumc\ initial conditions.

To further understand the general ``wrong sign" of $c^v_2\{4\}$ from hydrodynamic simulations with various initial conditions, we study the correlation between initial eccentricity $\varepsilon_2$ and final elliptic flow $v_2$. As shown in Fig.~\ref{fig:pv2e2} (a),  a clear deviation of elliptic flow from linear scaling is observed for $\varepsilon_2 > 0.5$ where the cubic term becomes significant, which is similar to the peripheral Pb--Pb collisions~\cite{Noronha-Hostler:2015dbi}~\footnote{It requires significant amount of computational resources to obtain Fig.~\ref{fig:pv2e2}.  We thus only show the results associated with  Para-III of \hijing\ initial conditions here.}. Such non-negligible cubic response leads to the fact that the scaled distribution $P(v_2/\langle v_2\rangle)$ and $P(\varepsilon_2/\langle\varepsilon_2\rangle)$ does not overlap with each other as shown in Fig.~\ref{fig:pv2e2} (b). It also introduces additional fluctuations of $v_2$ in the final states, which could even change the sign of $c_2^v\{4\}$ and make the model calculations fail to reproduce the negative $c_2^v\{4\}$ measured in experiments.

\begin{figure*}[t]
\begin{center}
\includegraphics[width=0.32\textwidth]{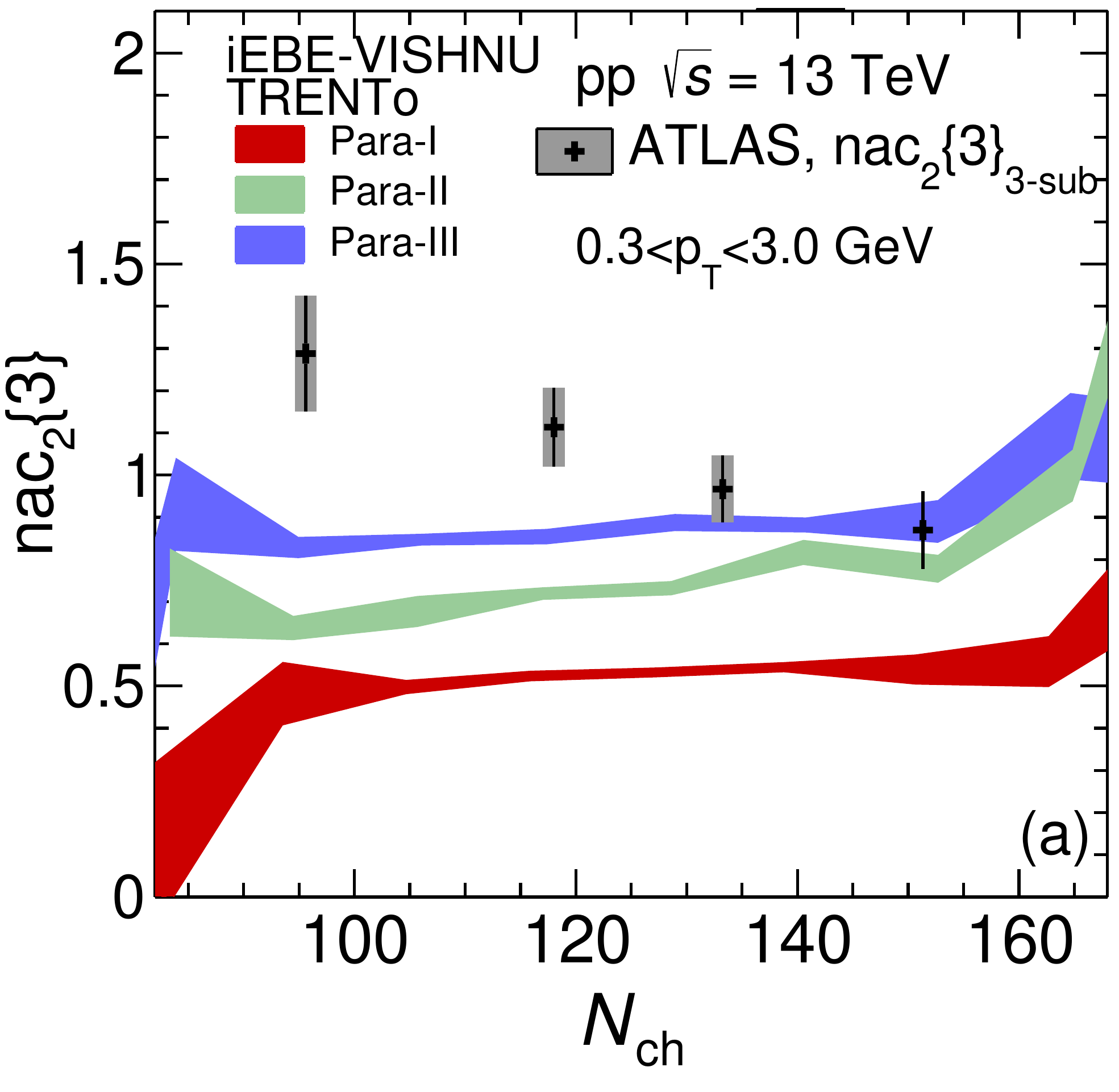}
\includegraphics[width=0.32\textwidth]{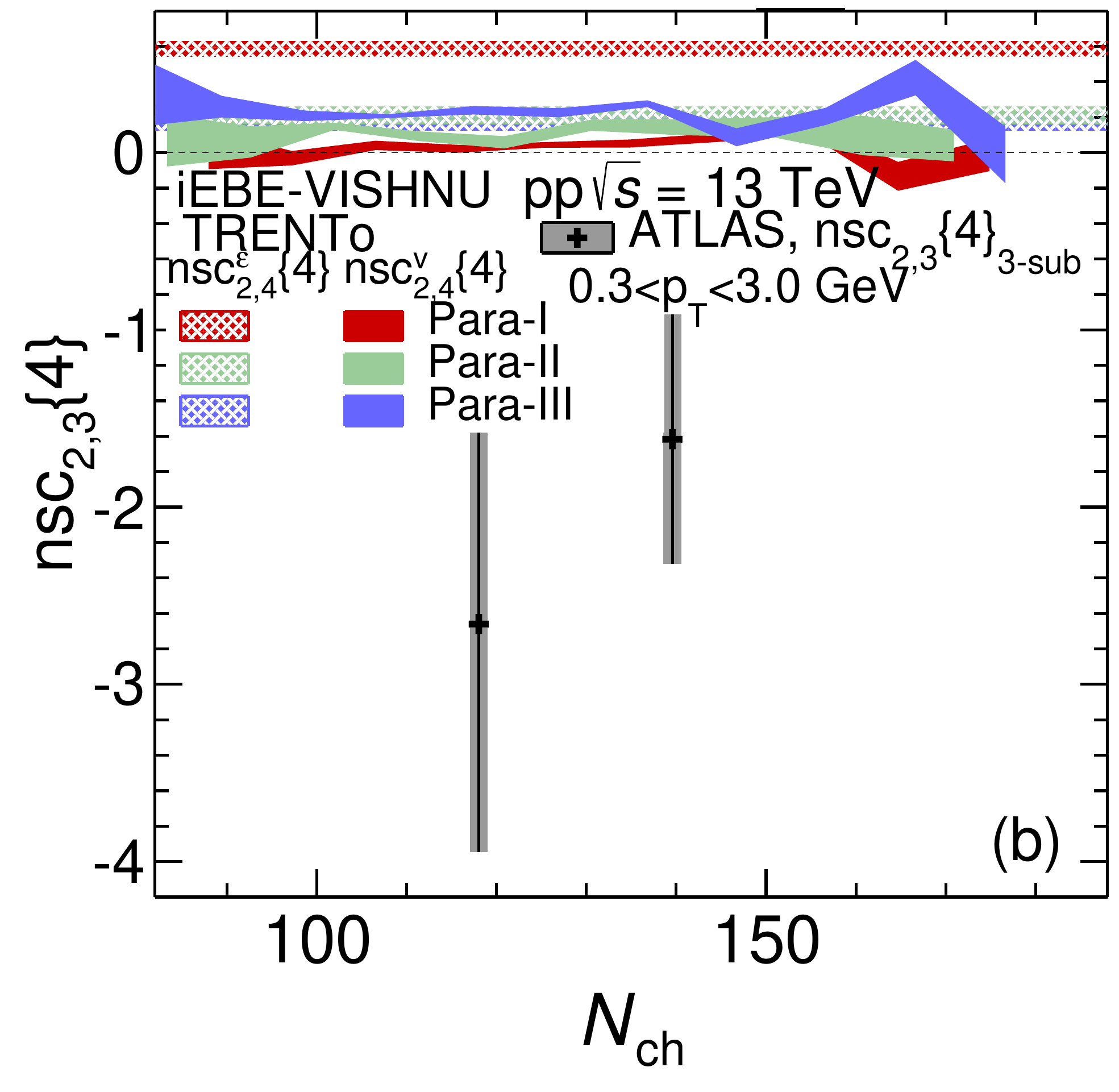}
\includegraphics[width=0.32\textwidth]{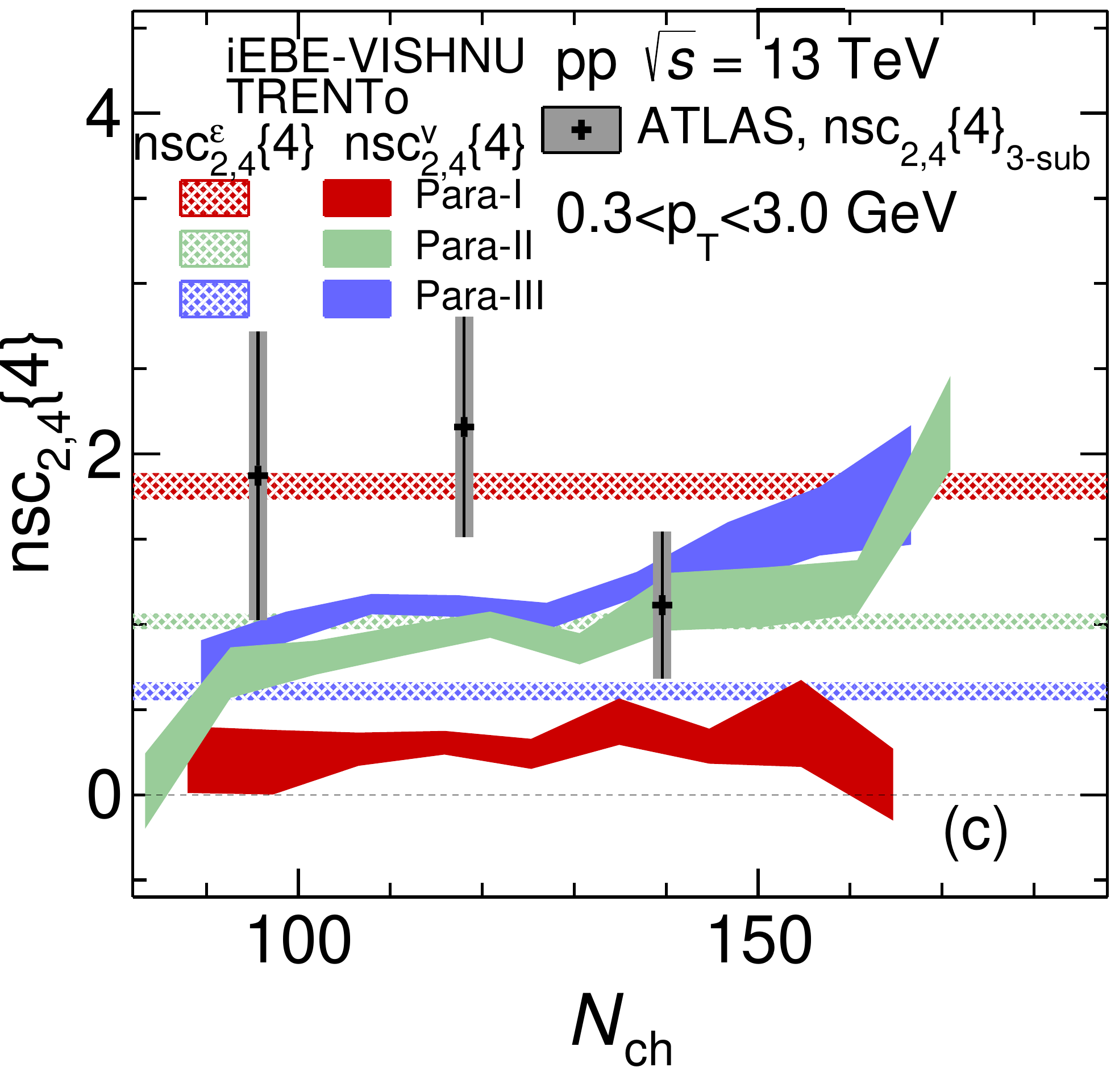}
\caption{(Color online)  $nac_{2}\{3\}$, $nsc_{2,3}\{4\}$  and $nsc_{2,4}\{4\}$   as a function of $N_{\rm ch}$ in p--p collisions at $\sqrt{s}=$  13 TeV, calculated by  \vishnu{} with \trento{} initial conditions, using standard cumulant method.   $nsc^\varepsilon_{2,3}\{4\}$ and $nsc^\varepsilon_{2,4}\{4\}$ of the  initial state in 0--0.1\% centrality bin are also shown.   The ATLAS data with three-subevent method are taken from Ref.~\cite{Aaboud:2018syf}.}
\label{fig:nac}
\end{center}
\end{figure*}

It has been generally argued that two- and multi-particle cumulants have different sensitivities to the flow fluctuations, which is written as~\cite{Voloshin:2008dg}
\begin{eqnarray}
v_{n}\{2\}^2 &=&  \langle v_n \rangle^2 + \sigma_v^{2}, \nonumber \\
v_{n}\{4\}^2 &=&  \langle v_n \rangle^2 - \sigma_v^{2}.
\label{eq1}
\end{eqnarray}
Here $\langle v_n \rangle$ and $\sigma_v$ represent the flow and flow fluctuations. These equations are valid in the case of small flow fluctuations, which might not be applied in small systems like p--p collisions. However, considering the fact that hydrodynamic calculations could quantitatively describe the two-particle correlations but could not even produce the correct sign of four-particle cumulants, one can conclude that the current hydrodynamic calculations could not simultaneously describe both the anisotropic flow $\langle v_n \rangle$ and the flow fluctuations $\sigma_v$.

In Fig.~\ref{fig:nac}, we further study the normalized  three- and four-particle azimuthal correlations in high-multiplicity proton--proton collisions at $\sqrt{s}=$ 13 TeV.  The three-particle asymmetric cumulant is defined as $ac_n\{3\} = \langle v_{n}^2 v_{2n} \cos 2n(\Psi_n-\Psi_{2n})\rangle $, which is sensitive to the correlations between flow magnitudes and the correlations between flow angles~\cite{Aad:2014lta,Acharya:2017zfg,Aaboud:2018syf}. The four-particle symmetric cumulants is defined as $sc_{m,n}\{4\} = \langle  v_{m}^2v_n^2 \rangle-\langle v_{m}^2 \rangle\langle v_{n}^2 \rangle $, which quantifies the correlation between $v_{m}^{2}$ and $v_{n}^{2}$~\cite{Bilandzic:2013kga,Zhu:2016puf}.
The corresponding normalized three- and four- particle cumulants are defined as~$nac_n\{3\}=ac_n\{3\}/(\langle v_{n}^2\rangle\cdot\sqrt{\langle v_{2n}^2 \rangle})$, $nsc_{m,n}\{4\} = sc_{m,n}\{4\}/(\langle v_{m}^2\rangle\cdot\langle v_n^2 \rangle)$, which try to eliminate the dependence on the flow coefficients and focus on evaluating the relative strength of the correlations between different flow harmonics.  Since the related  calculations are numerically expansive,  we only show the results $nac_n\{3\}$, $nsc_{2,3}\{4\}$  and $nsc_{2,4}\{4\}$ before and after hydrodynamic evolution with \trento{} initial condition. Fig.~\ref{fig:nac}  shows that $nac_n\{3\}$, $nsc_{2,3}\{4\}$ and $nsc_{2,4}\{4\}$  in the final states keep the same sign of those in the initial state correlations. Another interesting feature is that the hierarchy of the four-particle correlations in final states does not follow the one in the initial states. For example, Fig.~\ref{fig:nac} (b) shows that the $nsc_{2,3}^\varepsilon\{4\}$~\footnote{Here, the $nsc_{2,3}^\varepsilon\{4\}$ and $nsc_{2,4}^\varepsilon\{4\}$ are calculated in the 0--0.1\% centrality bin  in the initial states.  } from three sets of parameters follows Para-I  $>$ Para-II $>$ Para-III, but the hierarchy of the $nsc_{2,3}^v\{4\}$ is inverted after the hydrodynamic evolutions with Para-III  $>$ Para-II $>$ Para-I. This can be caused by the different non-linear response effects to various initial conditions. Such non-linear response effects are the greatest for Para-III, which lead to the largest  $nsc_{2,3}^v\{4\}$ after the hydrodynamic evolution. This is also consistent with the results of $c_2^v\{4\}$ in Fig.~\ref{fig:c24}, which shows that $c_2^v\{4\}$ of Para-III is the most positive one due to the largest non-linear response of $v_2$. \\

As shown in Figs.~\ref{fig:c24} and \ref{fig:nac}, the current hydrodynamic calculations with these three initial conditions have difficulties in describing the measured multi-particle single cumulants and mixed harmonic cumulants for high-multiplicity p--p collisions. Nevertheless the presented hydrodynamic calculations also confirm that the mixed harmonic multi-particle correlations are very sensitive to the details of initial conditions. If hydrodynamics works for the small p--p collision systems,  the related experimental data is very useful to constrain the corresponding initial conditions.

\section{Summary}\label{sec:summary}
In this paper, we investigated the hydrodynamic flow in high-multiplicity events of proton--proton collisions at $\sqrt{s}=$ 13 TeV, using \vishnu{} hybrid model with \hijing, \sumc{} and \trento{} initial conditions. With properly tuned parameters, \vishnu{} can roughly reproduce the measured two-particle correlations, including the integrated and differential flow for all charged and identified hadrons. However, the hydrodynamic calculations with any initial condition can not describe the negative $c_2\{4\}$ measured in experiments, which give a wrong sign. Further investigations showed that the elliptic flow $v_2$ does not linearly respond to the
initial eccentricity $\varepsilon_2$. The non-linear (cubic) response becomes important in the small systems, which
plays a non-negligible role and enhances the flow fluctuations. Such contribution
always leads to a positive $c^v_2\{4\}$ even when the sign of $c^\varepsilon_2\{4\}$ is negative in the initial conditions.

We also performed the first hydrodynamic calculations for normalized three- and four-particle azimuthal correlations, $nac_n\{3\}$, $nsc_{2,3}\{4\}$ and $nsc_{2,4}\{4\}$ in p--p collisions at $\sqrt{s}=$ 13 TeV, and found that \vishnu{} can qualitatively describe the features of $nac_n\{3\}$ and $nsc_{2,4}\{4\}$ but fail to reproduce the negative $nsc_{2,3}\{4\}$ as measured in experiments. At the current stage,  it is still challenging to describe the measured multi-particle cumulants of single and mixed harmonics within the framework of 2+1D hydrodynamics with these three initial conditions implemented in this paper. In the near future, it is worthwhile to implement 3+1D hydrodynamics with longitudinal fluctuations and dynamical initial conditions to further investigate these flow data in p--p collisions, which could help us to evaluate whether or not tiny droplets with collective expansion have been created in p--p collisions at the LHC.

\section{Acknowledgments}
We thank the discussions from C.~Bierlich, W.~Li, J.~Jia, B.~Schenke, J.~Schukraft and C.~Shun. WZ, KM and HS are supported by the NSFC under grant Nos.~11675004. Y.Z. is supported by the Danish Council for Independent Research, Natural Sciences, the Danish National Research Foundation (Danmarks Grundforskningsfond), the the Carlsberg Foundation (Carlsbergfondet) and a research grant (00025462) from VILLUM FONDEN. W.Z., K.M. and H.S. also gratefully acknowledge the extensive computing resources provided by the Super-computing Center of Chinese Academy of Science (SCCAS), Tianhe-1A from the National Supercomputing Center in Tianjin, China and the High-performance Computing Platform of Peking University.

%

\bibliography{bibliography}
%
\end{document}